\documentclass[reprint,showpacs,preprintnumbers,amsmath,amssymb, aip,superscriptaddress]{revtex4-1}
\usepackage{graphicx,wasysym}
\usepackage{dcolumn}
\usepackage{bm}
\usepackage{natbib}
\usepackage{tabulary}
\usepackage{hyperref}

\usepackage{color}
\usepackage[dvipsnames]{xcolor}
\usepackage[normalem]{ulem}

\newcommand{\kB}{k_\mathrm{B}}

\newcommand{\no}{\mathrm{N}_2\mathrm{O}_5}
\newcommand{\hno}{\mathrm{H N}\mathrm{O}_3}
\usepackage{bbold}

\begin{document}

\preprint{1}

\title{Elucidating the mechanism of reactive uptake of N$_2$O$_5$ in aqueous aerosol}
\author{Mirza Galib}
 \affiliation{Department of Chemistry, University of California}
\author{David T. Limmer}
 \email{dlimmer@berkeley.edu}
 \affiliation{Department of Chemistry, University of California}
\affiliation{Kavli Energy NanoScience Institute, Berkeley, California}
\affiliation{Materials Science Division, Lawrence Berkeley National Laboratory}
\affiliation{Chemical Science Division, Lawrence Berkeley National Laboratory}

\date{\today}

\maketitle

{\bf Nearly one third of all nitrogen oxides are removed from the atmosphere through the reactive uptake of $\no$ into aqueous aerosol. The primary step in reactive uptake is the rapid hydrolysis of $\no$, yet despite significant study, the mechanism and rate of this process are unknown.  Here we use machine learning-based reactive many body potentials and methods of importance sampling molecular dynamics simulations to study the solvation and subsequent hydrolysis of $\no$. We find that hydrolysis to nitric acid proceeds through the coordinated fluctuation of intramolecular charge separation and solvation, and its characteristic rate is 4.1 ns$^{-1}$, orders of magnitude faster than traditionally assumed. This large rate calls into question standard models of reactive uptake that envision local equilibration between the gas and the bulk solution. We propose an alternative model based on interfacial reactivity that can explain existing experimental observations and is corroborated by explicit simulations.
}

The heterogeneous hydrolysis of $\no$ plays a key role in establishing the oxidative power of the troposphere, and is a major factor in determining air quality and climate.\cite{sein_atmospheric_2016,crutzen_areps_1979}
In night time air, NO and NO$_2$ are oxidized by O$_3$ to form NO$_3$ and $\no$. \cite{brown_csr_2012}
Around 20\% of that atmospheric $\no$ is thought to be subsequently removed by hydrolysis to $\hno$ in aqueous aerosol.\cite{holmes2019role} 
However, a molecular level understanding of the reactive uptake of $\no$ is lacking,
frustrating attempts to rationalize variations in field measurements. \cite{davis_acp_2008,bertram_acpd_2009,abbatt_csr_2012,chang_ast_2011,mcduffie2018heterogeneous}
Using state of art computational tools, including machine learning based~\cite{singraber2019library,wang_cpc_2018} reactive force fields and methods of importance sampling molecular dynamics simulations, we have studied the reactive uptake in pure water. We have determined that the hydrolysis of $\no$ in aqueous aerosol is fast, occurring in less than 1 ns on average, and subsequently that interfacial processes dominate its reactive uptake. 
This finding is inconsistent with traditional models of reactive uptake, which assume reaction-limited bulk hydrolysis and equilibrium solvation.\cite{bertram_acpd_2009,chang_ast_2011,davidovits_cr_2006,poschl_ar_2011} Rather, we show with explicit simulations that reactive uptake can be understood as a result of competition between interfacial hydrolysis and evaporation. 

As an important reactive intermediate in the atmospheric chemistry of nitrogen oxides and nitrate aerosol, the heterogeneous chemistry of $\no$ has been the subject of intense study.\cite{mozurkewich_jgra_1988,bertram_acpd_2009,abbatt_csr_2012,chang_ast_2011,davidovits_cr_2006} Experimentally, only the overall mass transfer of $\no$ gas to aqueous aerosol can be easily measured, precluding a detailed understanding of the physical and chemical processes that underpin it.\cite{bertram_acpd_2009,davidovits_cr_2006,poschl_ar_2011} 
Under standard conditions, mass transfer is determined by the reactive uptake coefficient, $\gamma$, which is the fraction of $\no$ molecules that collide with an aerosol surface that are irreversibly lossed through reaction. Measurements of $\gamma$ in pure water vary between $0.01 \lesssim \gamma \lesssim 0.08$.\cite{bertram_acpd_2009,chang_ast_2011} Uptake on pure water aerosol represents a speed limit for typical atmospheric aerosol, as contributions from surface active organics  and soluble inorganic salts tend to suppress uptake.\cite{bertram_acpd_2009,park_jpca_2007,davis_acp_2008,ryder2015role} The size of $\gamma$ and its dependence on solution composition and thermodynamic state is currently rationalized with simplified kinetic models.\cite{chang_ast_2011,davidovits_cr_2006,akimoto_arc_2016} Unfortunately, the basic physical and chemical properties of $\no$, like its solubility and hydrolysis rate constant, that are needed to validate assumptions made in such models are not available.
Therefore a model capable of directly interrogating the molecular dynamics that transfer an initially gaseous $\no$ molecule into its solution hydrolysis products is needed.

Molecular simulations can in principle be used to 
gain microscopic insight into the reactive uptake of atmospheric gases into solution, but traditional theoretical methods are insufficient to reach the broad range of length and time scales required.\cite{davidovits_cr_2006} Classical force fields have been used to study the physical solvation of $\no$,\cite{li_jcp_2018,hirshberg_pccp_2018} where it is computationally tractable to employ enhanced sampling methods and represent large inhomogeneous systems. However, existing potentials are not suitable to model chemical reactions, precluding a study of the hydrolysis reaction.
\emph{Ab initio} molecular dynamics has been used to study hydrolysis and halide substitution reactions of $\no$ in water clusters.\cite{hammerich_pccp_2015,rossich2019microscopic,mccaslin_sa_2019,karimova_jpca_2019,mcnamara_pccp_2000} However, it is not typically feasible to study systems large enough to represent inhomogeneous systems or to evolve systems long enough to study rare events.  To overcome these limitations, we have employed machine learning techniques to fit a high dimensional reactive potential to \emph{ab initio} training data. The combination of novel potential representations and algorithms to fit them has recently enabled the use of machine learning based force fields for a range of complex chemical problems.\cite{singraber2019library,wang_cpc_2018} 
The resultant potential allows us to access larger length and time scales than typical \emph{ab initio} simulations, but with comparable accuracy.  In so doing we are able to employ advanced simulation methods to uncover a complete picture of the thermodynamics and reactive dynamics that lead to the uptake of $\no$. 

\section*{Results and Discussion}

In order to simulate the hydrolysis of $\no$ in liquid water, we have developed a reactive force field capable of describing a broad ensemble of solvation and bonding configurations. Specifically, we constructed a model using \emph{ab initio} reference data fit to a flexible artificial neural network functional.\cite{wang_cpc_2018} We used supervised and active learning procedures on a range of condensed phase and reactive path structures.\cite{SI} 
 The artificial neural networks are trained on reference energies and forces computed from density functional theory,\cite{zhang_prl_1998,grimme_jcp_2010} which provides an accurate description of aqueous solution structure and thermodynamics,\cite{bankura_jpcc_2014,galib_jcp_2017,morawietz_pnas_2016,marsalek_jcp_2017} and which we have additionally benchmarked for $\no$ gas phase dissociation energies.\cite{SI} All of our studies are at ambient conditions with temperature $T=$ 298K and pressure $p=1$ atm. The resultant reactive force field accurately represents the \emph{ab initio} potential-energy surface of water and $\no$, but at a significantly reduced computational cost, enabling the systematic study of the thermodynamics and kinetics of solvated $\no$ and its hydrolysis products.

\subsection*{Thermodynamics of solvation and hydrolysis.}

\begin{figure}[t]
\begin{center}
\includegraphics[width=8.5cm]{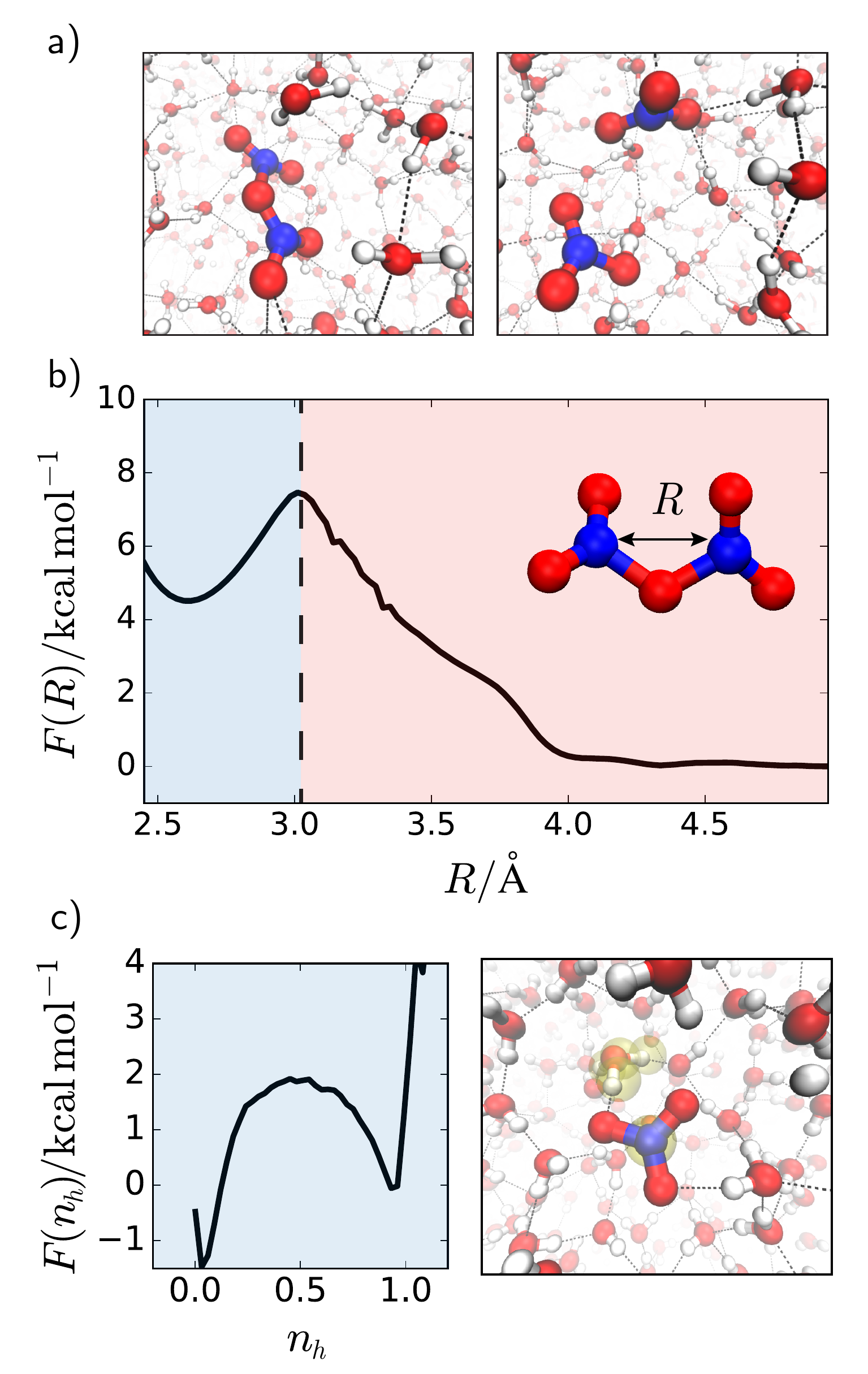}
\caption{Solvation and hydrolysis thermodynamics in bulk water. a) Representative snapshots of solvated $\no$ and $\hno$ in bulk water from molecular dynamics simulations. The red spheres denote oxygens, the blue nitrogens, and the white hydrogens. b) Free energy profile for $\no$ hydrolysis reaction as a function of intramolecular nitrogen-nitrogen distance. c) Free energy profile for the dissociation of $\hno$ as a function of a continuous coordination number between an O in the NO$_3$ moiety and a hydrogen, $n_h$. A characteristic snapshot of dissociated $\hno$ from molecular dynamics trajectory, where the excess proton is highlighted in yellow. }
\label{Fi:1}
\end{center} 
\end{figure}

Shown in Fig.~\ref{Fi:1}a) is a characteristic snapshot of $\no$ and its surrounding solvation environment generated from our neural network force field. The intramolecular structure of the solvated $\no$ is characterized by large fluctuations 
in the position of the center oxygen.\cite{SI} These fluctuations manifests the tendency of $\no$ to spontaneously undergo intramolecular charge separation,  localizing excess positive charge in an emergent NO$_2^{\delta+}$ moiety and excess negative charge in an NO$_3^{\delta-}$ moiety, as an transient precursor to dissociation.\cite{hirshberg_pccp_2018,mcnamara_pccp_2000} Despite the transient charge separation, we find that $\no$ is relatively weakly solvated on average. 
Water forms less than one hydrogen bond with the outer oxygens on average, and even fewer with the nitrogens and bridging oxygen, resulting in an unstructured solvation shell. This is because the localization of the charge is primarily on the nitrogens, which are typically sterically inaccessible.

The observed hydration structure is consistent with $\no$ being sparingly soluable in water. To quantify the driving force for dissolving $\no$ in water, we have computed the solvation free energy using thermodynamic perturbation theory.\cite{SI} 
The resultant solvation free energy, $\Delta F_s$, was determined to be  $\Delta F_s = -1.3 \pm 0.5$ kcal/mol implying a Henry's law constant 
of $H= 0.4\pm$ 0.1 M/atm. To our knowledge this is the first \emph{ab initio} estimate of the solubility of $\no$. It is much smaller than has been inferred from previous mass uptake experiments which range from 1-10 M.\cite{sander_acp_2015,mentel_pccp_1999} However, the interpretation of such experiments is difficult due to the inability to separate solvation of $\no$  from subsequent hydrolysis. For a molecule with a dipole this solubility is relatively low, though it is similar to other weakly solvated gases like $\mathrm{SO}_3$. The low solubility reflects a subtle interplay between favorable long range electrostatic energetics and a large unfavorable cavity formation entropy. 

The hydrolysis of $\no$ in liquid water is thermodynamically favorable. We have calculated the free energy for dissociating $\no$ using umbrella sampling. Specifically, we have computed the free energy as a function of the intramolecular nitrogen-nitrogen distance, $R$, as 
$
F(R) = -\kB T \ln \langle \delta (R -\hat{R} )\rangle 
$
where the hat denotes a fluctuating quantity, $\langle .. \rangle$ denotes a canonical ensemble average, $\kB$ is Boltzmann's constant, and $\delta$ is Dirac's delta function. The free energy is shown in Fig.~\ref{Fi:1}b), and exhibits a narrow minimum at $R=2.6  \mathrm{\AA}$ and a broad plateau for $R>4 \mathrm{\AA}$, separated by a barrier at $R=3 \mathrm{\AA}$.  The minimum at $R=2.6 \mathrm{\AA}$ reflects the intact $\no$ molecule, as shown in Fig.~\ref{Fi:1}a), while the plateau for $R>4 \mathrm{\AA}$ manifests its dissociation. 
We find that at large $R$ it is thermodynamically favorable to form two equivalents of $\hno$, also shown in Fig.~\ref{Fi:1}a). At relatively short separation distances, $4 \mathrm{\AA} \le R \le 6 \mathrm{\AA}$ only one of the two $\hno$ molecules are likely to be dissociated. The barrier region is wide, as large separations are needed to solvate the separated nitrogens. A barrier of nearly 4 kcal/mol implies that hydrolysis is a rare event, and that $\no$ can be dynamically distinguished from its eventual hydrolysis products. The free energy difference between the reactant and product basin is -4 kcal/mol. 
The low solubility of $\no$ implies that nearly all solvated $\no$ in pure water is transformed to $\hno$. 

After hydrolysis, it is thermodynamically favorable for the nascent nitric acid to dissociate into an excess proton and $\mathrm{NO}_3^-$. We have computed the free energy to deprotonate $\hno$ by monitoring a continuous coordination number, $n_h$, between the oxygens on the NO$_3$ moiety and a hydrogen.\cite{SI} The free energy, $F(n_h)$, can be estimated directly from
$
F(n_h) = -\kB T \ln \langle \delta (n_h - \hat{n}_h)\rangle
$, which is shown in Fig.~\ref{Fi:1}c). The free energy difference for removing a proton,  $F(n_h=0)-F(n_h=1)$ is -1.4 kcal/mol and corresponds to a pKa value of -1.1, which is reasonably close to the experimental value of -1.35.\cite{mckay_tfs_1956} 

 \begin{figure*}[t]
\begin{center}
\includegraphics[width=17cm]{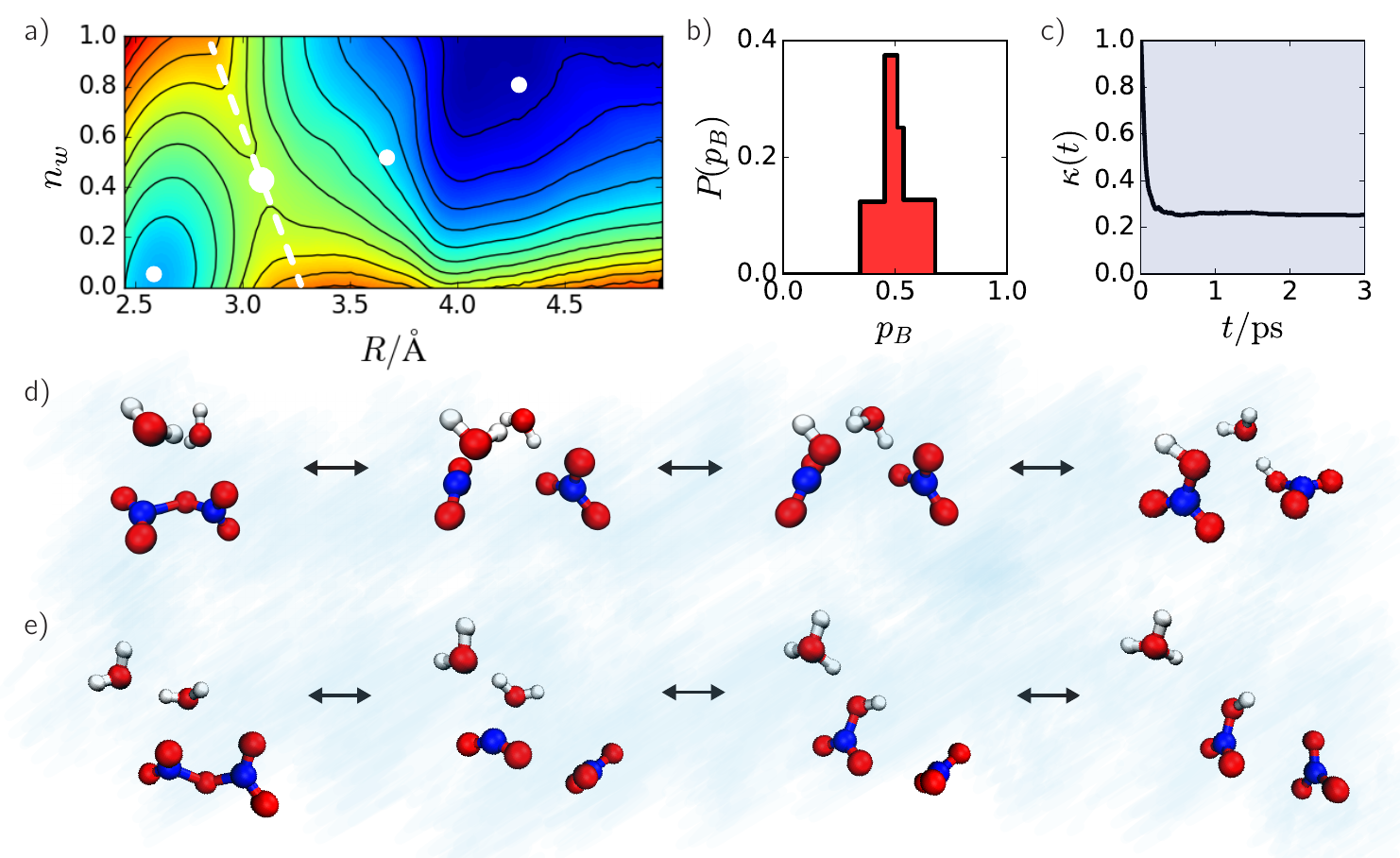}
\caption{Kinetics and microscopic reaction mechanisms for the hydrolysis of $\no$ in bulk water. a) Free energy as a function of the intramolecular nitrogen-nitrogen distance and a continuous water coordination number  $n_w$. Lines are spaced 1 kcal/mol apart and dashed lines plots the separatrix. The circles indicate the approximate location of configurations in d) and e). b) Distribution of commitment probabilities to the product basin, $p_B$, for configurations taken along the separatrix in a). c) Transmission coefficient, $\kappa(t)$, as a function of time. d) Representative snapshots along a  molecular dynamics trajectory in which two protonated nitric acids are formed through the concerted ionization of water and addition of the OH$^-$ to NO$_2^+$, followed by proton transfer to the NO$_3^-$.  e) Representative snapshots along a molecular dynamics trajectory in which hydrolysis of $\no$ into one $\hno$ and one NO3$^-$ proceeds through the ionization of water and addition of the OH$^-$ to the NO$_2^+$. }
\label{Fi:2}
\end{center} 
\end{figure*}

Taken together, the calculated thermodynamics of $\no$ solvation and subsequent hydrolysis in water are consistent with experimental observations that its accommodation into aqueous aerosol is largely irreversible.\cite{davidovits_cr_2006} 
Though weakly soluble, once in pure water $\no$ will undergo hydrolysis to form two $\hno$, which will subsequently deprotonate. 
Under high nitrate concentrations, or in low humidity droplets, this equilibrium could be shifted back towards an intact $\no$ and subsequently its reevaporation. Indeed, low water content droplets are observed to have smaller reactive uptake coefficients, and dissolved nitrate salts 
can reduce the reactive uptake by over an order of magnitude.\cite{mentel_pccp_1999,bertram_acpd_2009}
However, due to the effective irreversibility of the reaction, a complete understanding of reactive uptake requires insight into the kinetics of solvation and hydrolysis.

\subsection*{ Kinetics of $\no$ hydrolysis}
The mechanism of $\no$ hydrolysis involves an interplay between intramolecular charge separation and stabilization from the surrounding water. To understand this interplay, we identified a reaction coordinate that encodes the microscopic details relevant to hydrolysis in solution. An appropriate reaction coordinate is one that is capable of both distinguishing the intact $\no$ from its dissociation products, as well as characterizing the transition state ensemble of configurations, which are those configurations that have equal probability of committing to either the reactant or product states.\cite{geissler-jpcb_1999} While the nitrogen-nitrogen distance in Fig.~\ref{Fi:1}b) is capable of the former, it fails in the latter. Configurations taken at fixed values of $R$ are overwhelmingly committed to either the reactant or product basins of attractions.  This is because $R$ lacks direct information about the surrounding water, which is pivotal in describing hydrolysis. 

We have found that an appropriate reaction coordinate for hydrolysis is a linear combination of the nitrogen-nitrogen distance, $R$, and a continuous coordination number between the nitrogen atoms in $\no$ and the surrounding water molecules, denoted $n_w$.\cite{SI} Figure~\ref{Fi:2}a) shows the corresponding free energy surface, $F(n_w,R)$, computed from 
$
F(n_w,R) = -\kB T \ln \langle \delta (n_w - \hat{n}_w)\delta (R - \hat{R})\rangle \, 
$
using umbrella sampling.
The reactant basin with an $\no$ solvated in water and product basin are separated by the line $n_w = -3 R + 9.6$, which defines a separatrix distinguishing the two basins of attraction. The direction orthogonal to the separatrix we refer to as the reaction coordinate, $\xi$. For small $R$, the weak hydration structure of $\no$ is evident by the low value of $n_w$. The $\mathrm{NO}_3^{\delta -} + \mathrm{NO}_2^{\delta +}$ pair generated at large $R$ but $n_w=0$ are not thermodynamically stable. The hydrolysis products, two equivalents of $\hno$, at large $R$ have an elevated coordination number, $n_w=1$, reflecting the altered bonding arrangement upon abstracting a water molecule.  The saddle point of the surface, which we denote $\xi^{*}$, is located at an intermediate coordination number $n_w=0.4$, and intermediate nitrogen-nitrogen distance $R=3.1\, \mathrm{\AA}$, with a free energy barrier $\Delta F(\xi^{*}) =3.8$ kcal/mol. The thermodynamically most likely reactive path follows the simultaneous increase in the nitrogen-nitrogen distance and coordination number. The increasing distance correlates with the lengthening of a N--O bond and accompanying charge reorganization, which is thermodynamically stabilized by a solvent fluctuation that alters the coordination number. 
 
The correlated increase in $R$ and $n_w$ is not only thermodynamically favored, but also well characterizes the transition state ensemble for $\no$ hydrolysis. We have confirmed the latter by performing a committor analysis,\cite{pratt_jcp_1986,geissler-jpcb_1999} whereby the probability of configurations constrained to lie along the separatrix to commit to the product basin, $p_B$, is estimated by integrating an ensemble of trajectories from an initial Maxwell-Boltzmann distribution of velocities. If the dividing surface is a true representation of the transition state ensemble, there should be an equal probability to be committed to either reactant or product basins. Shown in Fig.~\ref{Fi:2}b) is the distribution of commitment probabilities for configurations taken along the separatrix. The distribution is peaked at $p_B = 0.5$ confirming that the combination of $n_w$ and $R$ is capable of characterizing the dynamics that lead to hydrolysis from $\no$.

We have employed the Bennett-Chandler method \cite{chandler_jcp_1978} to quantify the rate constant for hydrolysis. Specifically, we compute the rate, $k_\mathrm{h}(t)$, as a product of the transition state theory estimate, $k^\mathrm{TST}$, and the transmission coefficient, $\kappa(t)$, $k_\mathrm{h}(t)=\kappa(t) k^\mathrm{TST}$. The transition state theory estimate of the rate is computable from
$
k^\mathrm{TST}= \nu \exp[-\Delta F(\xi^{*})/\kB T]
$
where the prefactor $\nu$ is related to the mean velocity of $\xi$ in the reactant basin. The transmission coefficient corrects transition state theory for dynamical effects at the top of the barrier, and is given by the plateau region of the flux-side correlation function.\cite{SI}
The transmission coefficient is shown in Fig.~\ref{Fi:2}c) and plateaus to a value of 0.25 within 0.5 ps. 
Taken together we find the rate of hydrolysis to be $k_\mathrm{h}=4.1 \,\mathrm{ns}^{-1}$, implying an average lifetime of $\no$ to be nearly $240$ ps. This time is in excellent agreement with that estimated from 100 individual reactive trajectories propagated with direct dynamics. 

Figures~\ref{Fi:2}d and \ref{Fi:2}e) show representative snapshots taken along hydrolysis pathways generated from our molecular dynamics trajectories. Subsequent to passing through the transition state, we find that the ensemble of reactive pathways bifurcate resulting in two different product states.  In one pathway, Fig~\ref{Fi:2}d), two nitric acids are formed 
through the concerted ionization of water and addition of the OH$^-$ to the NO$_2^+$ moiety, followed by proton transfer to NO$_3^-$. In the other pathway, Fig~\ref{Fi:2}e),
one $\hno$ and one NO$_3^{-}$ are formed. As in the first pathway, this process proceeds through the ionization of water and addition of the OH$^-$ to the NO$_2^+$, however, the $\mathrm{H_3O}^+$ generated does not have an existing hydrogen bond wire to enable the subsequent donation of the proton to the NO$_3^{-}$. In our ensemble of 100 trajectories, 20 \% of those follow the first pathway and 80 \% follow the latter one. These product distributions and the pathways that evolve them are similar to previous calculations in water clusters.\cite{rossich2019microscopic}   
During hydrolysis, we find NO$_2^+$ is only formed transiently, with an average lifetime of 4 ps, and is better characterized by a hydrated $\mathrm{H_2 O NO_2}^{+}$ species than a stable intermediate. Once an $\hno$ molecule has its own independent solvation shell, we find that dissociation occurs on average within 60 ps, or that ionization to NO$_3^-$ + H$_3$O$^+$ occurs with a reaction rate of 15.4 ns$^{-1}$. The Grotthuss diffusion of the excess protons are well reproduced with our force field. We find a relative diffusivity of $\mathrm{H_3O}^+$ to $\mathrm{OH}^-$ of 2.2 compared to the 1.9 measured experimentally.\cite{mills_els_2013}

Previous estimates of the hydrolysis rate for $\no$ in solution place it on the order of 10$^{-4}\,\mathrm{ns}^{-1}$, or four orders of magnitude slower than our computed rate.\cite{chang_ast_2011} However, like the solubility of $\no$, this rate has been inferred indirectly from mass transfer models and not measured independently. The model most commonly invoked assumes equilibration between the vapor and bulk solution, and is valid when uptake is reaction limited.\cite{davidovits_cr_2006} Given the short lifetime of $\no$  in solution, this equilibrium assumption requires reevaluation. 
Indeed, the reaction diffusion length, defined as $\ell_\mathrm{D} = \sqrt{D/k_\mathrm{h}}$ where $D$ is the self-diffusion constant of $\no$
is small. Within our model $\ell_\mathrm{D}\approx 1 $nm. 
 As a consequence, we expect that $\no$ does not diffuse away from the interface before reacting. This suggests that rather than being mediated by bulk solvation and subsequent reaction, reactive uptake of $\no$ is determined directly at the air-water interface, through a process of interfacial absorption and reaction. A model for reactive uptake based on such interfacial activity is expanded on below. 

\begin{figure*}[t]
\begin{center}
\includegraphics[width=17cm]{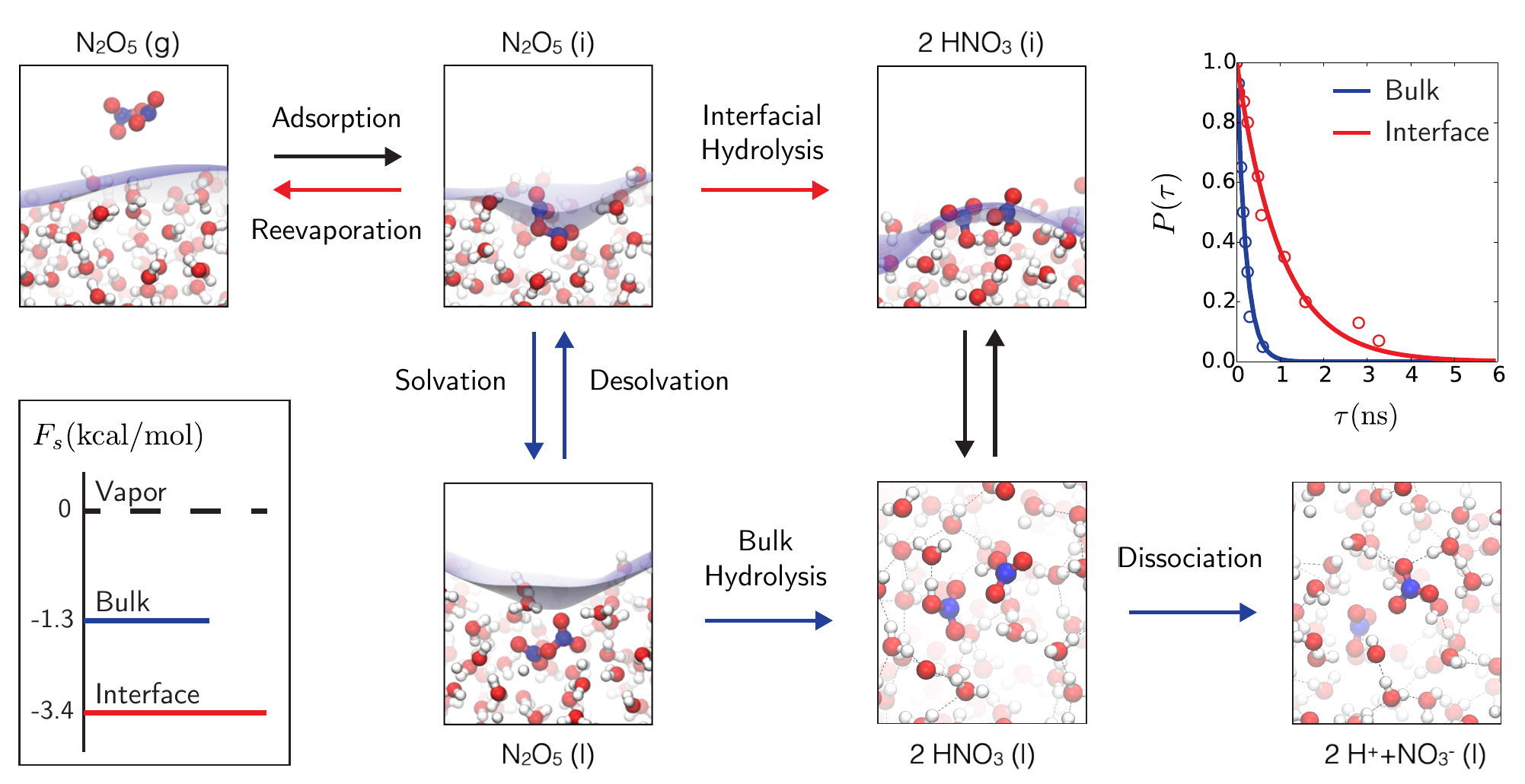}
\caption{Elementary physical and chemical steps involved in the reactive uptake of $\no$ in a pure water droplet. The red arrows denote our proposed interfacial model, while blue refer to the standard bulk model. The panels show an incoming $\no$ molecule first adsorbed at the liquid-vapor interface, which then either reacts to form $\hno$ or evaporates back into the gas phase. Diffusion into the bulk of the droplet is comparatively slow, but once in the bulk, $\no$ can undergo hydrolysis. In either case, deprotonation of $\hno$ occurs after solvation into the bulk. The blue surface in the top panels represents the location of the liquid-vapor interface. Top right panel: Probability distribution of the lifetime time of a $\no$ molecule, $P(\tau)$, at the air-water interface and in bulk water. Bottom left panel: Solvation free energies of $\no$ at the air-water interface and in bulk compared to the gaseous $\no$.}
\label{Fi:3}
\end{center} 
\end{figure*}

\subsection*{Interfacial model for N$_2$O$_5$ reactive uptake}

The canonical kinetic model for the reactive uptake of $\no$ is the so-called resistor model.\cite{worsnop1989temperature,akimoto_arc_2016,davidovits_cr_2006,poschl_ar_2011}
This model assumes that the gas molecule is first accommodated at the surface, with probability $\alpha$, and then diffuses from the surface to the bulk where the reaction takes place. The bulk reaction with rate $k_\mathrm{h}$, should be slow enough that an equilibrium can be established between the gas and the liquid phase, with concentrations determined by the Henry's law constant $H$. Under these assumptions for the mass transfer kinetics, the reactive uptake coefficient, $\gamma$, can be estimated from 
\begin{equation}
\gamma=\left ( \dfrac{1}{\alpha} + \dfrac{v }{4 \kB T H\sqrt{D k_\mathrm{h}}} \right )^{-1}
\end{equation}
where $v$ is the thermal velocity. Measurements suggest a value of $\alpha \gtrsim 0.4$,\cite{grvzinic2017efficient} however as discussed above, neither $H$ nor $k_\mathrm{h}$ can be independently measured. Previous work has assumed a value of $H$ to be 5.0 M/atm, taken from extrapolating the known solubilities of a series of other $\mathrm{NO}_x$ compounds.\cite{mentel_pccp_1999} Inverting the expression for $\gamma$ and setting it equal to the 0.03, which is the middle of the range of experimental estimates, provides an estimate of the reaction rate on the order of 10$^5$ s$^{-1}$ for $\no$ hydrolysis. This analysis is internally consistent, as it predicts a reaction-diffusion length much larger than the width of the interface, $\ell_\mathrm{D}\approx 80 $nm, but the solubility and hydrolysis rate are dramatically different from those computed \emph{ab initio}. Using our computed values of $H$ and $k_\mathrm{h} $, we arrive at $\gamma=0.6$, much higher than observed.

This inconsistency can be resolved by formulating an alternative to the standard resistor model that envisions the reactive uptake of $\no$ as an interfacial process. Specifically, assuming all incoming $\no$ stick to the interface and do not diffuse away, the reactive uptake is given by a competition between hydrolysis of $\no$ at an interface\cite{rossich2019microscopic} and its reevaporation back to the gas phase. If $k^\mathrm{s}_\mathrm{h}$ is the reaction rate at the surface and $k_\mathrm{e}$ is the evaporation rate, then the reactive uptake coefficient can be computed from
\begin{equation}
\gamma= \frac{k^\mathrm{s}_\mathrm{h}}{k^\mathrm{s}_\mathrm{h}+k_\mathrm{e}}
\end{equation}
which in the limit that $\gamma$ is small reduces to $\gamma \approx k^\mathrm{s}_\mathrm{h}/ k_\mathrm{e}$.\cite{hanson1997surface} This competition is illustrated in Fig.~\ref{Fi:3} with accompanying simulation snapshots, and contrasts it with the processes of solvation and bulk hydrolysis included in the standard resistor model. This interfacial model is analogous to an older perspective on $\no$ uptake from Mozurkewich and Calvert\cite{mozurkewich_jgra_1988}. Using molecular dynamics simulations, we have tested the assumptions of this model and explicitly computed $\gamma$ by estimating $k^\mathrm{s}_\mathrm{h}$ and $k_\mathrm{e}$. 

We have computed the reaction rate at the air-water interface to be $k^\mathrm{s}_\mathrm{h}=0.95$ ns$^{-1}$ from direct molecular dynamics simulations. The distribution of waiting times for hydrolysis for both the interface and bulk are shown in Fig.~\ref{Fi:3}. This rate is slower than the corresponding rate in the bulk by a factor of 4, and predominantly follows a pathway that generates two protonated $\hno$ molecules. This is consistent with previous reports of the weaker acidity of $\hno$ at the air-water interface.\cite{shamay2007water} 
We have estimated the evaporation rate by first computing the free energy of adsorption to the interface from the vapor using thermodynamic perturbation theory and then assuming that evaporation is barrierless. We obtained a free energy of adsorption of $\Delta F_\mathrm{s} =-3.4$ kcal/mol, which is lower than the corresponding solvation free energy, as shown in Fig.~\ref{Fi:3}.  This indicates that $\no$ is preferentially solvated at the interface, which is consistent with previous studies using empirical potentials\cite{hirshberg_pccp_2018,li_jcp_2018} and the weak hydration observed in our bulk simulations. From this, we estimate an evaporation rate of $k_\mathrm{e}= 12.5$ ns$^{-1}$.  The reactive uptake coefficient computable from these two rate processes yields $\gamma= 0.07$, which is in reasonable agreement with the experimental range.\cite{bertram_acpd_2009,chang_ast_2011} 
 
An interfacial model of $\no$ reactive uptake helps rationalize a number of existing experimental observations, and opens new questions for further examination. For example, it has been noted that the temperature dependence of $\no$ uptake is rather weak.\cite{van_jpc_1990} The similar barrier heights for interfacial hydrolysis and evaporation result in both processes increasing with temperature at about the same rate, leaving $\gamma$ nearly invariant in our model. Further, measurements of the reactive uptake on ice particles are close to those for liquid particles.\cite{apodaca_acp_2008} The importance of surface processes elucidated in our work clarifies this coincidence, as diffusion into the bulk of the solid is prohibitively slow, and as we have shown hydrolysis can still proceed. 
Finally, the rapid rate of hydrolysis observed here explains why the uptake coefficient does not strongly depend on reactions with inorganic species in solution\cite{bertram_acpd_2009}  as such reactions cannot kinetically compete for intact $\no$.
However, it is known that the branching ratio for $\no$ decomposition in solutions with halide anions, $\mathrm{X}^{-}$ and excess nitrate strongly favors $\mathrm{XNO}_2$ over $\hno$, beginning at 1 M $\mathrm{X}^{-}$ concentration.\cite{sobyra_jpca_2019}
This seems at odds with the rapid hydrolysis to $\hno$, which reacts with $\mathrm{X}^{-}$ to form $\mathrm{XNO}_2$ only at very low pH. Surface chemistries not viable in the bulk solution such as those  catalyzed by enhanced interfacial proton concentrations could be studied to clarify this. Such studies are now possible by employing analogous neural network based simulations as we have developed here. With these tools, many heterogeneous chemistries previously defying explanation can now be systematically studied and understood.

\section*{Methods}

\noindent {\bf Machine learning \emph{ab initio} potential.} We have used the DeePMD-kit\cite{wang_cpc_2018} to learn the many body interatomic potential energy and forces generated at the DFT level of theory.  The primary data sets for the training were generated from ab initio molecular dynamics simulations using the Gaussian Plane Wave(GPW) implementation in CP2K.\cite{vandevondele_cpc_2005} All \emph{ab initio} molecular dynamics simulations were carried out in the canonical  ensemble at ambient temperature and density using the revised version of PBE functional\cite{zhang_prl_1998} along with empirical dispersion correction (Grimme D3)\cite{grimme_jcp_2010}. We used a molopt-DZVP basis set and a plane wave cut-off of 300 Ry. The core electrons were described with GTH pseuodopotential.\cite{goedecker_prb_1996} We also carried out metadynamics simulations\cite{laio_rpp_2008} to generate reactive structures along the hydrolysis pathway. With the primary data set generated by molecular dynamics and metadynamics simulations, we first trained two independent machine learned potentials that were then followed by active learning to improve both models. The disagreement in force between the two models was used to select the new data sets for active learning.  Final convergences for testing errors in the energy were 0.2 meV/atom. 

\vspace{.5cm}
\noindent {\bf Molecular dynamics simulations.} To investigate the hydrolysis reaction of $\no$, we carried out molecular dynamics simulations at ambient temperature and pressure with 0.5 fs timestep. The integrator used a Langevin thermostat, with characteristic time constant of 1 ps. The bulk system contained one $\no$ molecule solvated by 253 water molecules in a 19.73 x 19.73 x 19.73 \AA~ box with periodic boundary conditions in all three dimensions. An equilibration molecular dynamics simulations of 5 ns was carried out by classical molecular dynamics that was followed by another equilibration MD simulations for 400 ps with machine learned force field. During the equilibration, the N-N distance was constrained to ~2.6 \AA~ to prevent the hydrolysis reaction taking place. We sampled the initial configurations from a 1 ns constrained molecular dynamics simulation and then carried out unconstrained molecular dynamics simulations for an ensemble of 50 trajectories, each for 1 ns.

\noindent To investigate the hydrolysis reaction of $\no$ at the air-water interface, we prepared a slab model with thickness of 25 x 25 x 25 \AA, having free interface and an additional 20  \AA~ vacuum on each side. We employed periodic boundary conditions in all three dimensions. The slab model included 1 $\no$ molecule and 522 water molecules. The $z$-position of the $\no$ molecule was constrained at the Gibbs diving surface of the slab. The initial configuration was generated from an equilibrated water box. An equilibration of 10 ns was carried out by classical molecular dynamics simulations with SPC/E water and GAFF force field,\cite{hirshberg_pccp_2018} which was then followed by another equilibration MD simulations for 500 ps with the machine learned force field. During the equilibration, the N-N distance was constrained to ~2.6 \AA~ to prevent the hydrolysis reaction taking place. We sampled the initial configurations from a 1 ns constrained molecular dynamics simulation and carried out molecular dynamics simulations for an ensemble of 28 trajectories, each for 3 ns. 

\noindent {\bf Free energy and rate calculations.} We used umbrella sampling \cite{frenkel2001understanding} to estimate the reaction free energies for the hydrolysis reaction of $\no$ in the bulk water. Harmonic biases were employed for both $n_w$ and $R$, and each of 26 windows were simulated for 1 ns. The free energies were then estimated using WHAM.\cite{kumar1992weighted} In order to calculate the correction to the transition state theory rate we computed the transmission coefficient from an ensemble of 2000 unbiased trajectories. The commitor probability was computed from an ensemble of 1000 unbiased trajectories, each starting from constrained configuration at the dividing surface with a random velocity taken from Maxwell-Boltzmann distribution.
We have computed the free energy to dissociated $\hno$ in bulk water from an ensemble of  molecular dynamics trajectories having one $\hno$ molecule solvated by 255 water molecule in a 19.73 x 19.73 x 19.73 \AA ~box. Since deprotonation occurs frequently, we have computed the free energy by monitoring $n_h$.
We have computed the solvation free energy of $\no$ using thermodynamic perturbation theory. For computational efficiency, we first used an empirical nonreactive reference potential and constructed a reversible work path by pulling a molecule of $\no$ initially in the vapor through a liquid-vapor interface and into the bulk using a slab geometry. We used the SPC/E water model and a GAFF force field for the $\no$ with partial charges parametrized to reproduce the \emph{ab initio} electrostatic potential.\cite{hirshberg_pccp_2018} We then estimated the free energy difference between the empirical model and our neural network potential model, by linearizing the relative Boltzmann weights collected from 20,000 configurations of the solvated classical model. An analogous calculation was used to compute the absorption free energy at the interface. 
\\

\noindent {\bf Acknowledgments}
The authors thank Timothy Bertram, Benny Gerber, Andreas Goetz and Gilbert Nathanson for stimulating discussions and Barak Hirshberg for initial solvated $\no$ configurations. This work was funded by the National Science Foundation through the National Science Foundation Center for Aerosol
Impacts on Chemistry of the Environment (NSF-CAICE) under Grant No. CHE 1801971. This research used resources of the National Energy Research Scientific Computing Center (NERSC), a U.S. DOE Office of Science User Facility operated under Contract No. DE-AC02-05CH11231.

%\bibliography{bib}


% Produces the bibliography via BibTeX.
\end{document}

\noindent{\bf Supplemental Information: \\
Elucidating the mechanism of reactive uptake of N$_2$O$_5$ in aqueous aerosol}

\section{ Gas phase dissociation energy of $\no$} We have computed the dissociation energy difference and the energy barrier for the gas phase dissociation reaction ($\no$ $\rightarrow$ NO$_3^-$ + NO$_2^+$) at the revPBE-D3 level of theory\cite{zhang_prl_1998,grimme_jcp_2010} and compared those to B3LYP \cite{becke98density,lee1988development} and MP2\cite{moller1934note}. These are shown in Table~\ref{Table:En}, in which we find that revPBE functional can satisfactorily reproduce the gas phase dissociation energies relative to a hybrid functional and higher level electronic structure theory.

\begin{table}[h!]
\label{Table:En}
  \begin{center}
    \caption{Calculated energies (kcal/mol) for the gas phase dissociation of $\no$ into NO$_3^-$ and NO$_2^+$}
    \begin{tabular}{l|c|c} 
      \hline
       \textbf{Method} & \textbf{$\Delta$E} & \textbf{$\Delta$E{$^\dagger$}}\\
      \hline
       MP2/6311++G(2d,2p)$^*$& 147.33 & 12.21\\
       B3LYP/6311++G(2d,2p)$^*$ & 155.85 & 14.00\\
       revPBE/6311++G(2d,2p)$^*$& 151.74 & 13.78\\
      \hline
    \end{tabular}
  \end{center}
\end{table}

\section{Machine learning ab initio potentials}

\subsection{Training Data set} The training data set includes structures generated by \emph{ab initio} molecular dynamics with three dimensional periodic boundary conditions for solvated $\no$ and bulk ambient water. We have included representative structures from the following simulations in a 19.73 x 19.73 x 19.73 \AA~ simulation box:
\begin{enumerate}
\setlength{\parskip}{0pt} 
\setlength{\itemsep}{0pt plus 1pt}
\item pure water box with 256 water molecules 
\item solvated $\no$ with one $\no$ molecule solvated by 253 water molecules 
\item solvated $\hno$ with one $\hno$ molecule solvated by 253 water molecules 
\item solvated NO$_3^-$ and H$_3$O$^+$ with one NO$_3^-$ and one H$_3$O$^+$ molecules solvated by 252 water molecules 
\item solvated H$_3$O$^+$ and OH$^-$ with one H$_3$O$^+$ and one OH$^-$ molecules solvated by 254 water molecules  
\item solvated $\no$ at the surface of a liquid vapor interface with one $\no$ and 522 water molecules in a 25 x 25 x 25 \AA~ slab and 20 \AA~ vacuum on both sides. 
\end{enumerate}
We also included structures along the hydrolysis pathway into our training sampled from metadynamics\cite{laio_rpp_2008} simulations in conjunction with AIMD using CP2K. Our final data set to train the bulk solvated $\no$ model contained 10000 data points and the slab model contained 20000 data points.

\subsection{Training accuracy}

\begin{figure}[h!]
\begin{center}
\includegraphics[width=8.5cm]{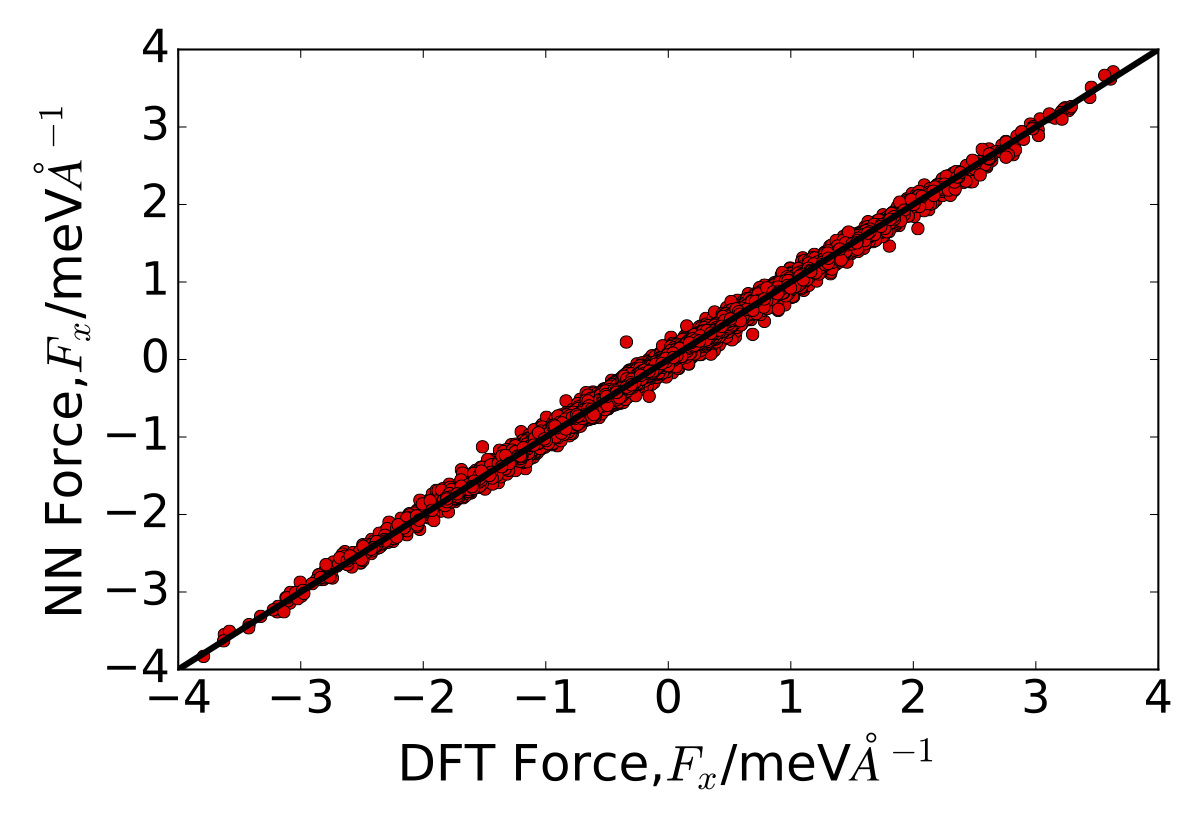}
\caption{ NN model prediction vs DFT results for the x-component of the force over a data set that was included in the training. }
\label{Fi:1}
\end{center} 
\end{figure}

\begin{figure}[h!]
\begin{center}
\includegraphics[width=8.5cm]{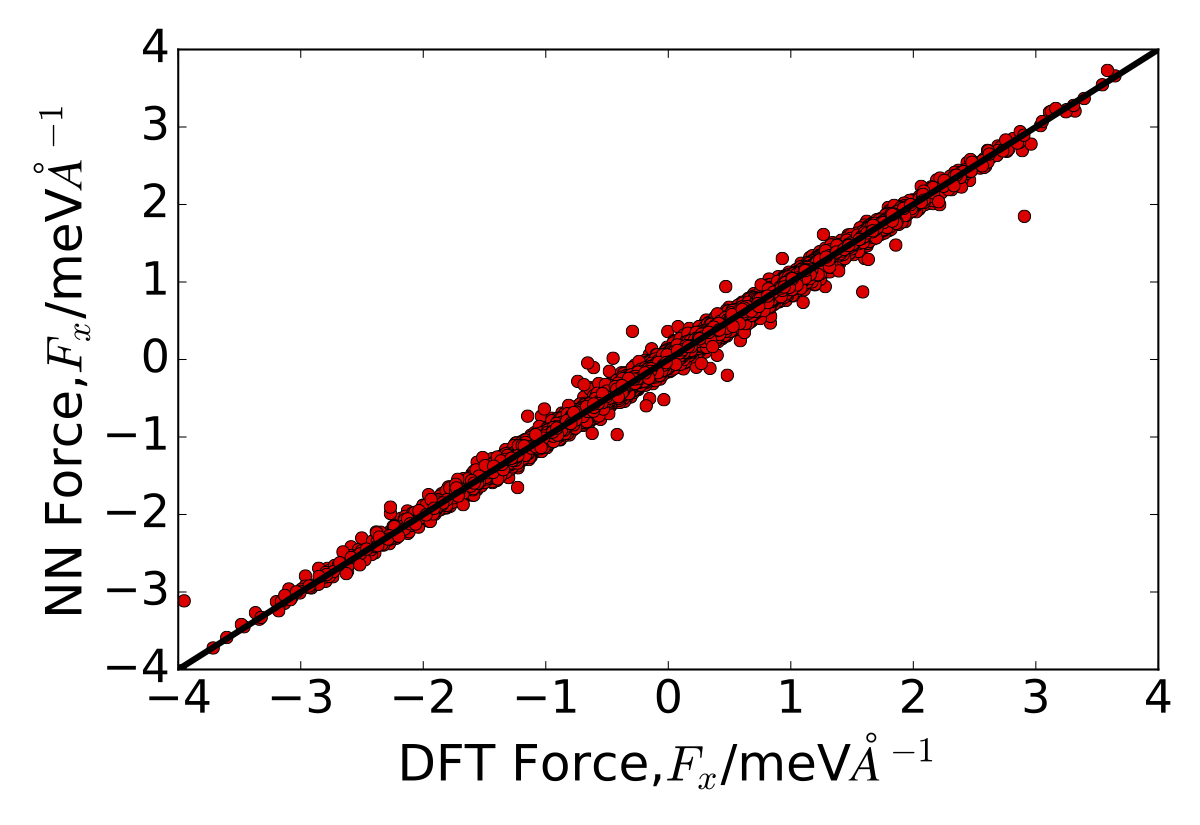}
\caption{NN model prediction vs DFT results for the x-component of the force over a data set that was not included in the training.}
\label{Fi:2}
\end{center} 
\end{figure}

We trained our neural network (NN) model using a deep neural network architecture with 3 layers each having 600 nodes as implemented in DeePMD-kit.\cite{wang_cpc_2018} We used a cut off distance of 8 \AA \, to represent the local internal structure around any atom in our training data set. We defined the loss function as the sum of mean square deviations in energy and force, which was minimized during the training process. For the slab model, we also included mean square deviation in virial in the loss function. At the end of the training, the accuracy of the model was 2.0 meV in energy/atom and 0.05 eV/\AA~ in force/atom. Correlation plots for the testing and training data for the force predicted by the neural network relative to that computed from DFT are shown in Fig.\ref{Fi:1} and \ref{Fi:2}. Summary errors are presented in Table \ref{tab:table1} and Table \ref{tab:table2}. 

\begin{table}[h!]
  \begin{center}
    \caption{Bulk Model}
    \label{tab:table1}
    \begin{tabular}{l|c|c} 
      \hline
       \textbf{} & \textbf{Training data set} & \textbf{Test data set}\\
      \hline
      Energy (meV/atom) & 0.2 & 0.25\\
      Force (meV/\AA-atom) & 50 & 60\\
      \hline
    \end{tabular}
  \end{center}
\end{table}

\begin{table}[h!]
  \begin{center}
    \caption{Slab Model}
    \label{tab:table2}
    \begin{tabular}{l|c|r} 
      \hline     
      \textbf{} & \textbf{Training data set} & \textbf{Test data set}\\
      \hline
      Energy (meV/atom) & 0.2 & 0.25\\
      Force (meV/\AA-atom) & 60 & 70\\
      Virial (meV/atom) & 3.0 & 3.0 \\
      \hline
    \end{tabular}
  \end{center}
\end{table}

\clearpage
\subsection{Active learning} 
\begin{figure}[h!]
\begin{center}
\includegraphics[width=8.5cm]{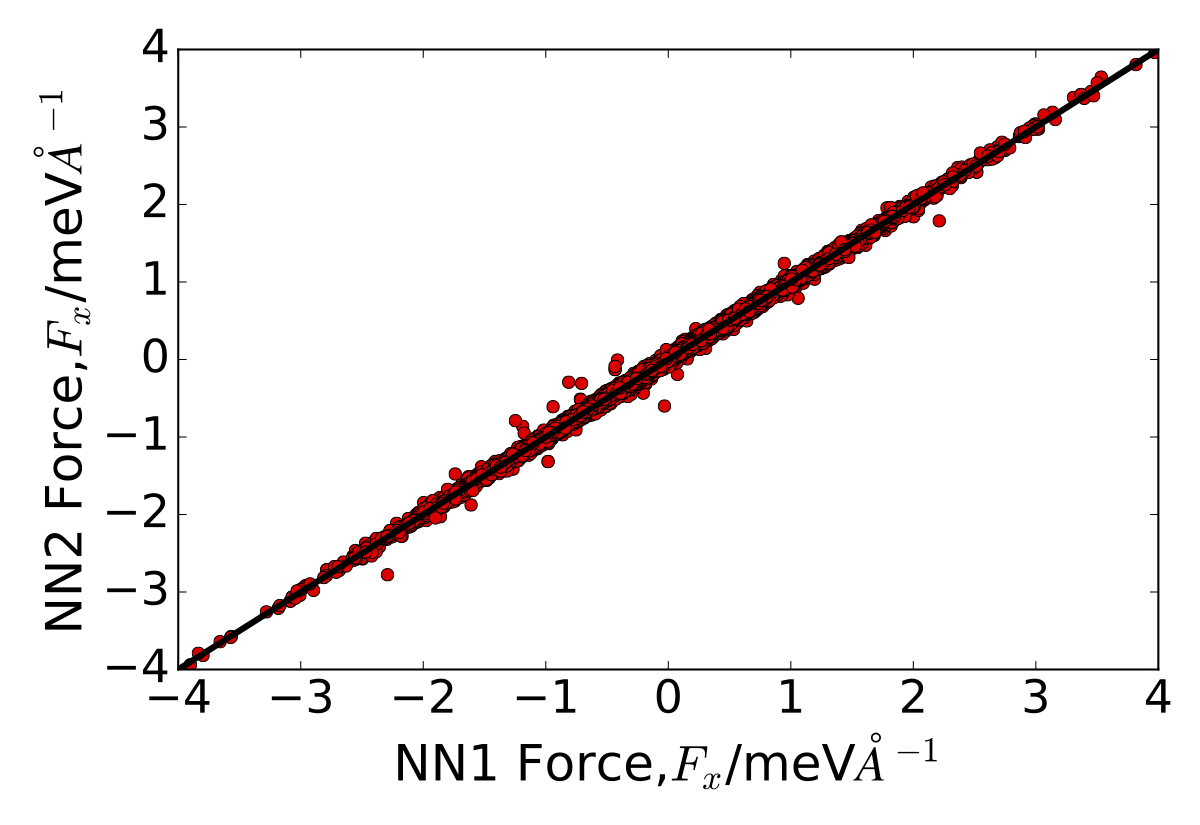}
\caption{NN model prediction between two independently trained models for the x-component of the force over a data set that was not included in the training.}
\label{Fi:3}
\end{center} 
\end{figure}

We trained two independent models, each with the same primary data set but with different initial parameters and different structures of the hidden layers. Both models were trained with a neural network architecture having 3 layers, but one having 600 nodes in each layer and the other having 400 nodes in each layer. We then followed  active learning procedure to improve both models.\cite{behler2015constructing} The disagreement in force between the two models was used to select the new data sets for active learning. The force-force correlation between these two model is used to determine finally whether the training data set is large enough to achieve a converged result. This data is shown in Fig.~\ref{Fi:3}, and yielded a mean squared deviation of 45 meV/\AA-atom.

\subsection{Validation of the NN model}

\begin{figure}[h!]
\begin{center}
\includegraphics[width=8.5cm]{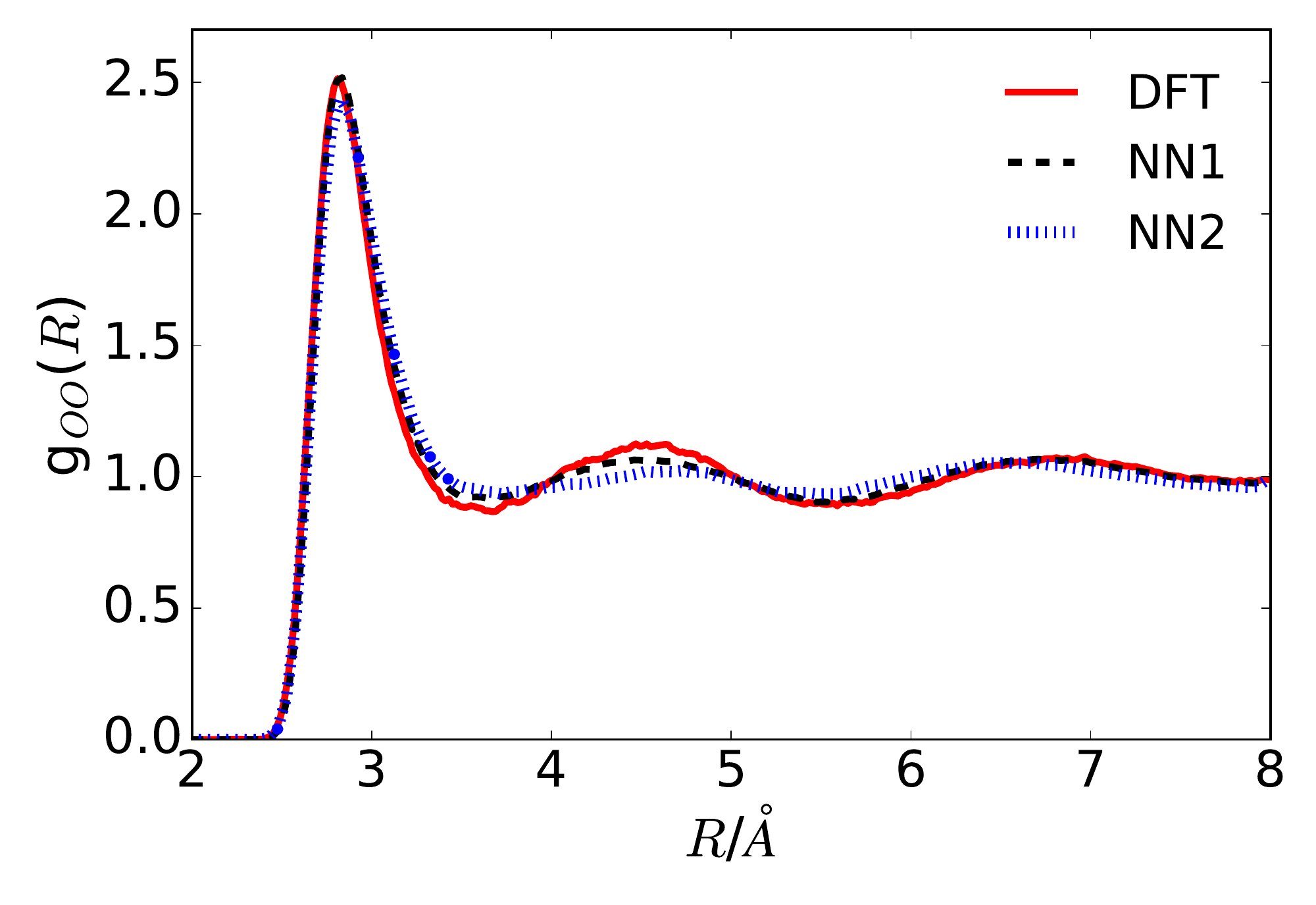}
\caption{O-O radial distribution function for bulk water at ambient temperature and pressure}
\label{Fi:4}
\end{center} 
\end{figure}

\begin{figure}[h!]
\begin{center}
\includegraphics[width=8.5cm]{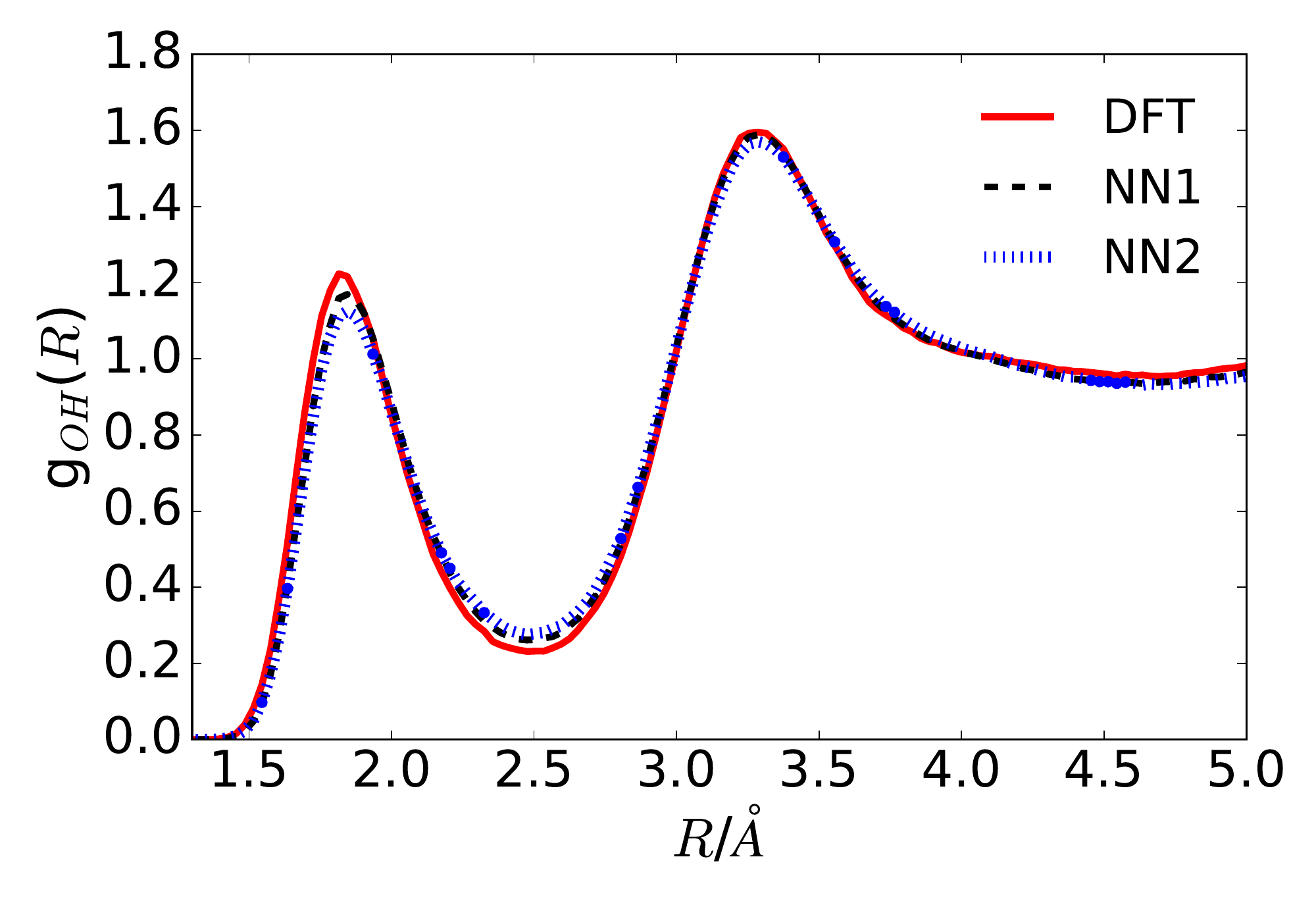}
\caption{O-H radial distribution function for bulk water at ambient temperature and pressure}
\label{Fi:5}
\end{center} 
\end{figure}

\begin{figure}[h!]
\begin{center}
\includegraphics[width=8.5cm]{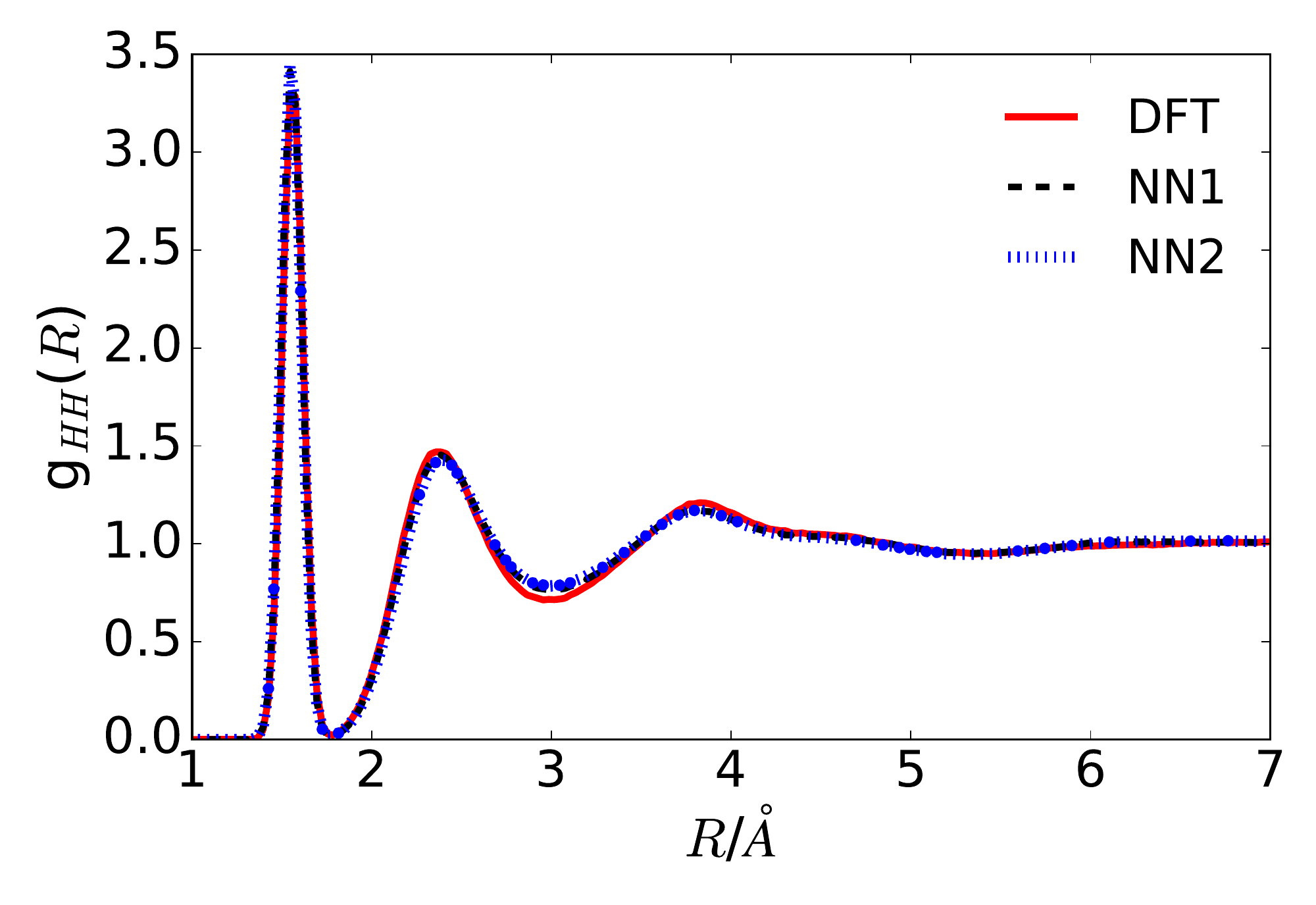}
\caption{H-H radial distribution function for bulk water at ambient temperature and pressure}
\label{Fi:6}
\end{center} 
\end{figure}

\begin{figure}[ht!]
\begin{center}
\includegraphics[width=8.5cm]{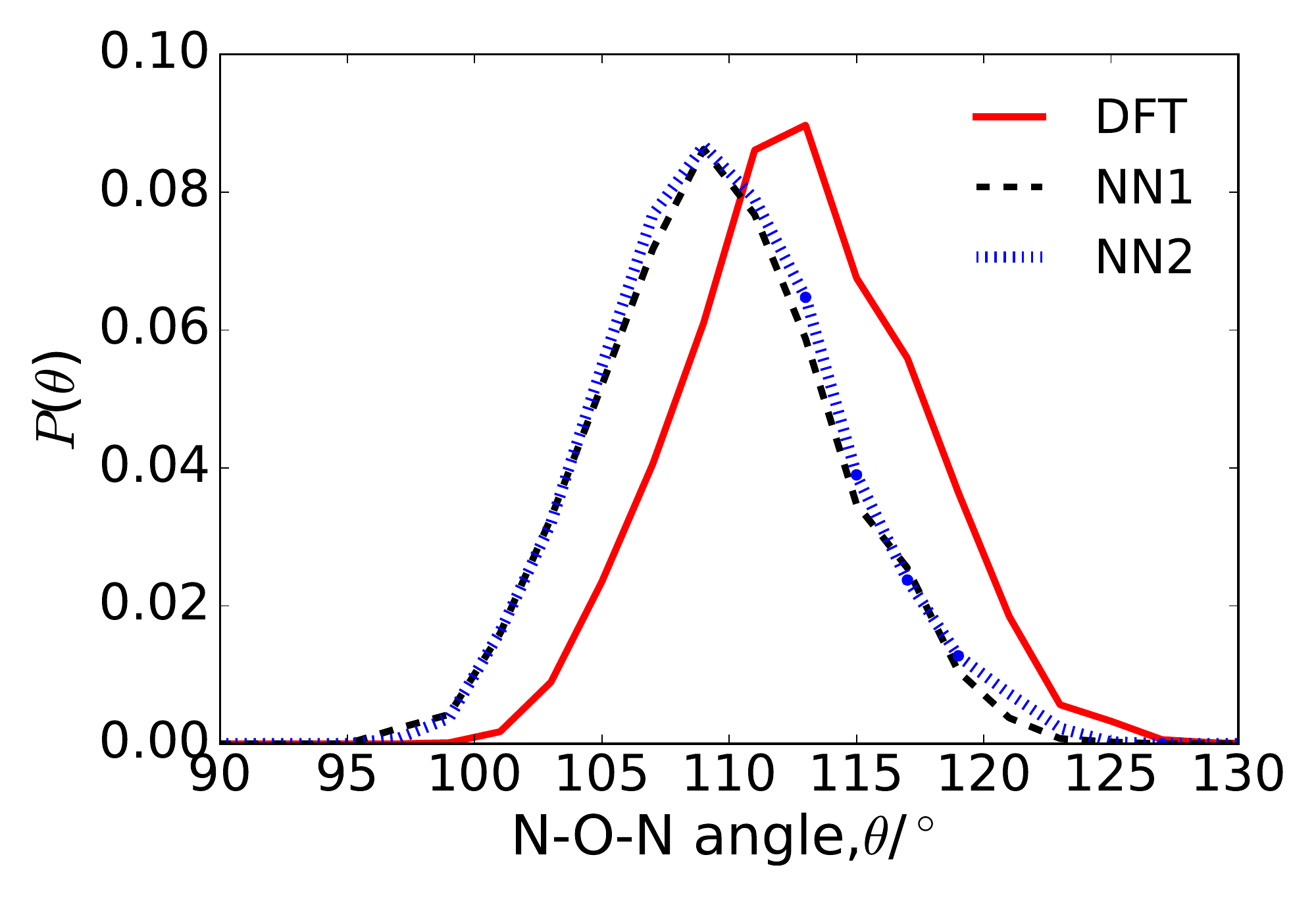}
\caption{Probability distribution of N-O-N angle of solvated $\no$ in bulk ambient water}
\label{Fi:7}
\end{center} 
\end{figure}

\begin{figure}[h!]
\begin{center}
\includegraphics[width=8.5cm]{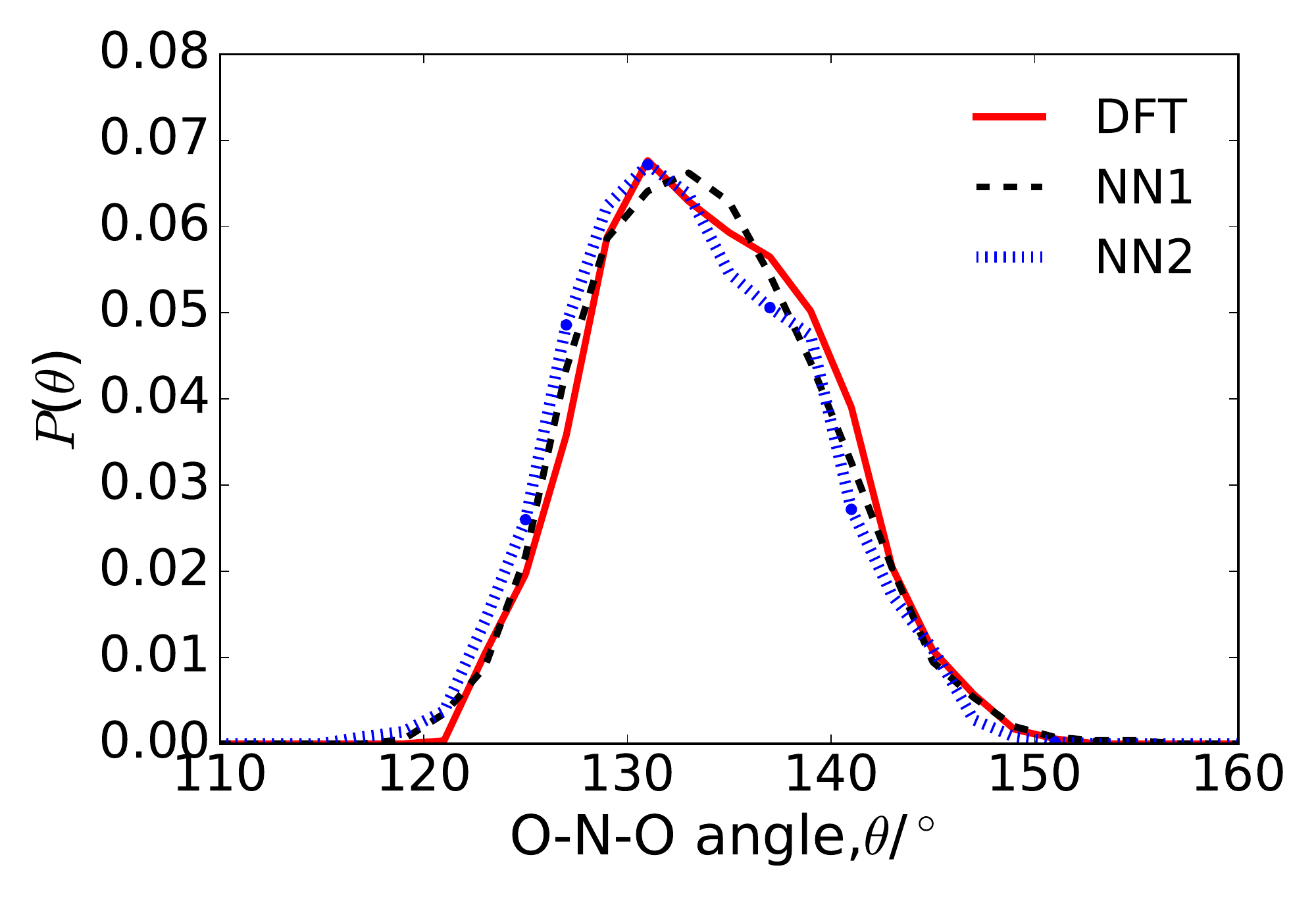}
\caption{Probability distribution of the terminal O-N-O angles of solvated $\no$ in bulk ambient water}
\label{Fi:8}
\end{center} 
\end{figure}

\begin{figure}[h!]
\begin{center}
\includegraphics[width=8.5cm]{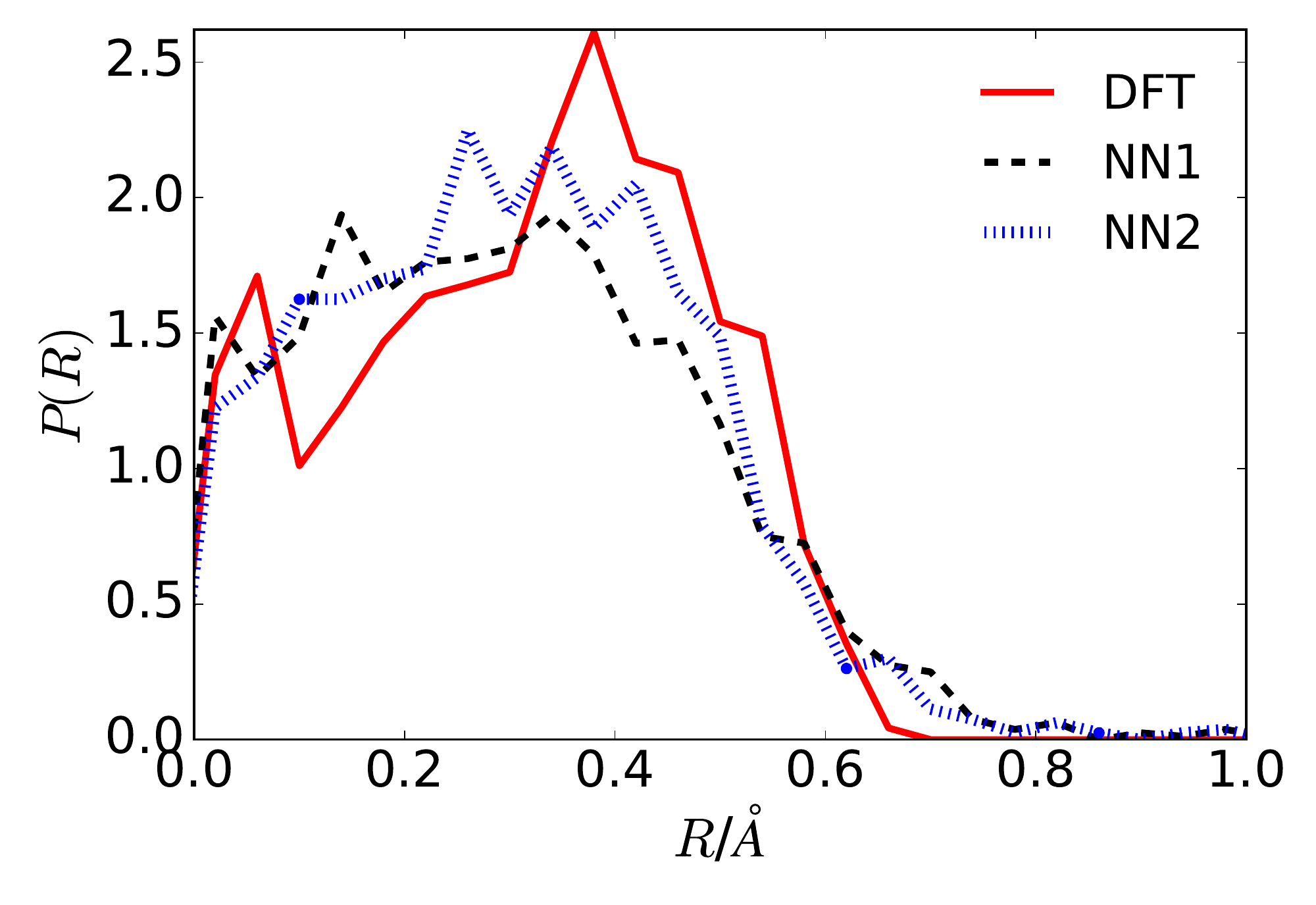}
\caption{Probability distribution of the difference between two N-O distances of solvated $\no$ in bulk ambient water}
\label{Fi:9}
\end{center} 
\end{figure}

\begin{figure}[h!]
\begin{center}
\includegraphics[width=8.5cm]{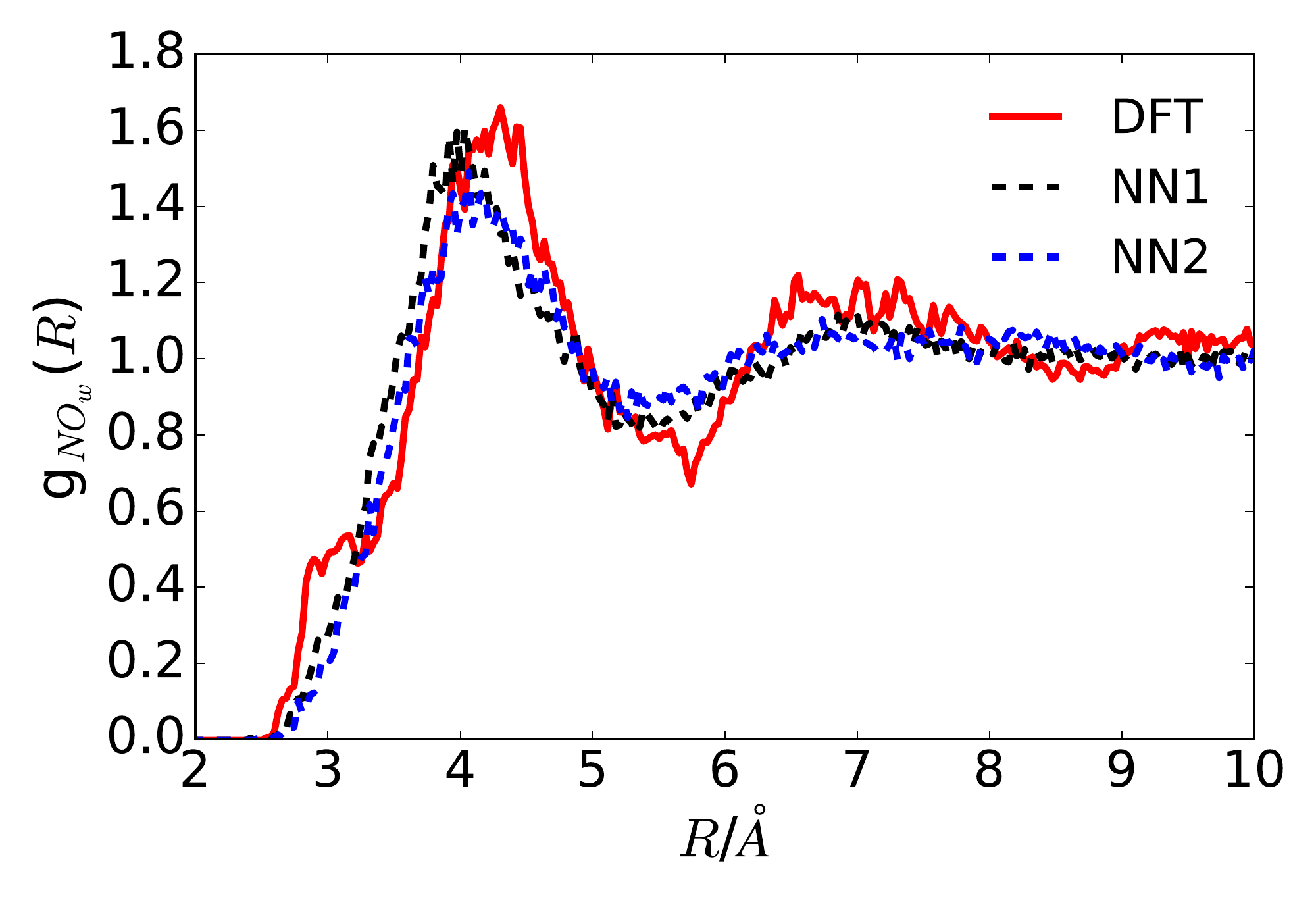}
\caption{Nitrogen-water oxygen radial distribution function for solvated $\no$ in bulk water at ambient temperature and pressure}
\label{Fi:10}
\end{center} 
\end{figure}

We have checked the accuracy of our model by comparing various structural and dynamic properties with DFT data and also the consistency between the two independent NN models. We report the structural calculations in Figs.\ref{Fi:4}-\ref{Fi:10}.  We have calculated the diffusion constant of $\no$, H$_3$O$^+$ and OH$^-$ in bulk water from the mean squared displacement of the solvated molecule in a molecular dynamics trajectory.  This is detailed in Table~\ref{Tab:Diff}.

\begin{table}[h!]
  \begin{center}
    \caption{Diffusion constant, units are in x 10$^{-9}$ m$^2$/s }
    \label{Tab:Diff}
    \begin{tabular}{l|c|c|c} 
      \hline
      \textbf{} & \textbf{Simulation} & \textbf{Experiment\cite{mills_els_2013,anttila_jpca_2006}}\\
      \hline
      $\no$ & 0.55  & 0.10 \\
       H$_3$O$^+$& 7.8  & 9.4\\
       OH$^-$  & 3.6 &  5.2 \\
      \hline
    \end{tabular}
  \end{center}
\end{table}

\newpage
\section{Free energy calculations}

\subsection{Free energy of $\no$ hydrolysis} We employed umbrella sampling to calculate the free energy of $\no$ hydrolysis, as a function of two reaction coordinates, i) the distance between the two nitrogen atoms within the $\no$ molecule ($R$) and ii) the water coordination number $n_w$. The coordination number was computed from\cite{iannuzzi2003efficient}
\begin{equation}
 n_w = \sum_{\substack{i \in {O_{\no}}\\ j \in H_{\mathrm{H_2O}}}} \frac{1-(\frac{r_{ij}}{r_c})^{4}}{1-(\frac{r_{ij}}{r_c})^{16}}
 \end{equation} 
 where $r_{ij}$ is the distance between atoms $i$ and $j$, and $r_c=2.4$ \AA. We employed harmonic potentials of the form,
 $$
U_B = k_R \left (R-R_0 \right )^2 + k_n \left (n_w-n_0 \right )^2
 $$
 with 26 windows are equally spaced along the distance coordinate,  $2.4 \mathrm{\AA} \le R_0 \le 5.0 \mathrm{\AA} $ and 10 windows along the coordination number coordinate, $0\le n_0 \le 1.0$. We used spring constants of $k_R=k_n=$15.0 kcal/mol-\AA$^2$. Each of the windows were run for 1 ns. 

\subsection{$\hno$ dissociation} To compute the free energy for dissociation of $\hno$, we monitored a continuous coordination number, $n_h$, between the oxygens on the NO$_3$ moiety and the hydrogens, defined as
 \begin{equation}
 n_h = \sum_{\substack{i \in {O_{\mathrm{NO_3}}}\\ j \in \mathrm{H}}} \frac{1-(\frac{r_{ij}}{r_c})^{12}}{1-(\frac{r_{ij}}{r_c})^{24}}
 \end{equation} 
with $r_c= 1.2$\AA. Using an ensemble of 10 trajectories, each for 4 ns we were able to converge a distribution of $n_h$ and also the characteristic time for of $\hno$.

\subsection{Bulk and interfacial solvation free energy of $\no$ }
\begin{figure}[h!]
\begin{center}
\includegraphics[width=8.5cm]{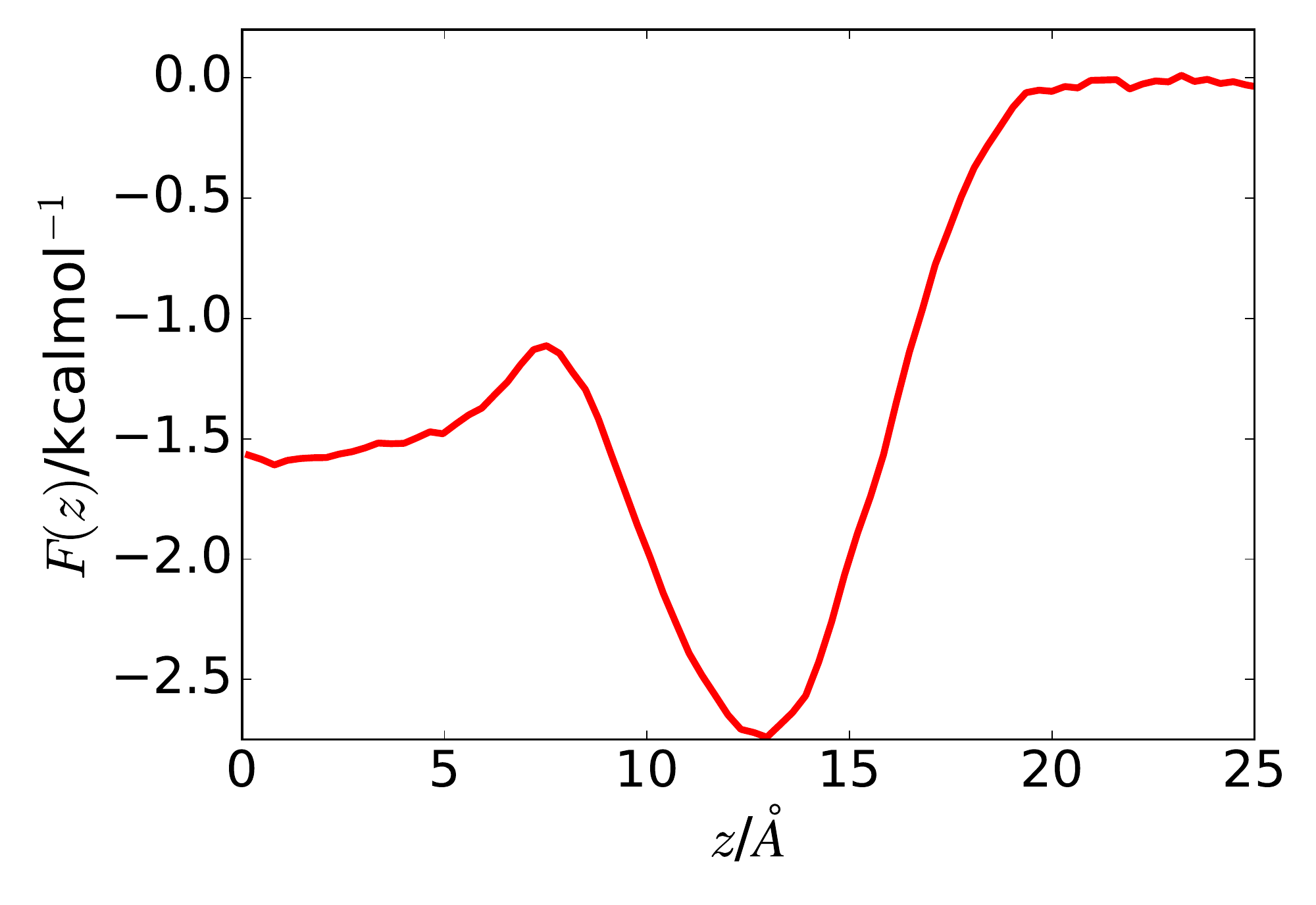}
\caption{Potential of mean force for the transfer of a $\no$ molecule from gas phase to bulk of the liquid. Gibbs dividing surface is at 12 \AA~ and gas phase is in the right hand side of the dividing surface.}
\label{Fi:13}
\end{center} 
\end{figure}

In order to computer the free energy of solvation, we employed thermodynamic perturbation theory using a classical fixed charge reference potential. 
We have computed the free energy profile for the transfer of a $\no$ molecule from the gas phase to the bulk through the air-water interface in slab simulations using umbrella sampling\cite{frenkel2001understanding}. We employed SPC/E water model\cite{berendsen1987missing} and the GAFF force field\cite{wang_jcc_2004}  for the $\no$ with partial charges parametrized to reproduce the \emph{ab initio} electrostatic potential\cite{hirshberg_pccp_2018}. We employed a real-space cutoff of 9.0 \AA~ to non-bonded interactions and the long range electrostatics was computed by Particle Mesh Ewald summation. The bonds involving hydrogen atoms were constrained using the SHAKE algorithm.\cite{miyamoto1992settle} The temperature was kept at 300 K using Langevin dynamics with a collision frequency of 5.0 ps$^{-1}$ and a time step of 1 fs was used as employed by LAMMPS.\cite{plimpton1995fast} The free energy profile is obtained from a set of umbrella sampling using distance between the COM of a water slab and the $\no$ molecule. The bias potential was of the form
$$
U_B = k_z \left (z-z_o \right )^2
$$
where we took 50 windows equally spaced along the z-coordinate from 0.0 $\le z_o \le$ 25.0 \AA, and employed $k_z=$5.0 kcal/mol-\AA$^2$. Each of the windows were run for 4 ns. The free energy as a function of $z$, $F(z)$, is shown in Fig.~\ref{Fi:13}. We define a bulk $\no$ as $z<4\AA$ and an interface  $\no$  as $10\le z \le 17$\AA.

To compute the solvation free energy within the NNP from the empirical model, it is sufficient to compute the additional free energy to transform the gaseous $\no$ from the NNP from the empirical model and the solvated $\no$ NNP from the empirical model. 
The free energy to transform the system from the empirical model described by energy function $E_\mathrm{E}$to the NN model described by energy function $E_\mathrm{NNP}$ is given exactly by,\cite{frenkel2001understanding}
$$
F_\mathrm{NNP} - F_\mathrm{E} = -\kB T \ln \langle e^{-\beta (E_\mathrm{NNP}-E_\mathrm{E})}\rangle_\mathrm{E}
$$  
where $F_\mathrm{NNP}, F_\mathrm{E}$ are the energy energy in the NN model and empirical model respective, and $\langle ... \rangle$ denotes ensemble average. If the energy difference between the two representations is small, the expression above can be expanded to first order, yielding,
$$
F_\mathrm{NNP} - F_\mathrm{E} \approx  \langle  E_\mathrm{NNP}\rangle -\langle E_\mathrm{E}\rangle_\mathrm{E}
$$  
which is an approximation that can be check \emph{a posteriori}. For the solvation free energy in the bulk and at the interface, we must compute free energy changes for the gasoes $\no$ as well as the two differently solvated species. 
 For the empirical model, we have computed the average energies for these three species from a 10 ns long molecular dynamics trajectory, and  averaged potential energy from a 4 ns long molecular dynamics trajectory from the NN model. 
These differences, as well as their implied free energies are shown in Tables \ref{TabF} and \ref{TabFF}. Note that the first order perturbation theory assumes identical entropic contribution to the free energy, which given the small changes in energy, on the order of $\kB T$, seem self-consistent.

 \begin{table}

  \begin{center}
    \caption{ Enthalpy and entropic contribution to the solvation free energy computed from the empirical potential. Units are in kcal/mol. The statistical sampling error of the calculated free energy values ($\Delta F$) is approximately $\pm$ 0.1 kcal/mol and that for the energetic contribution ($\Delta U$) is $\pm$ 1.0 kcal/mol.}
     \label{TabF}
    \begin{tabular}{l|c|c|c} 
    \hline
      \textbf{} & \textbf{$\Delta F$} & \textbf{$\Delta U$} & \textbf{-T$\Delta S$}\\
      \hline
      Gas & 0.0 & 0.0 & 0.0 \\
      Surface & -2.7 & -7.5 & 4.8 \\
      Bulk & -1.55 & -9.2 & 7.65 \\
      \hline
    \end{tabular}
  \end{center}
\end{table}
 
  \begin{table}

  \begin{center}
    \caption{ Solvation free energy  obtained from the NN model employing thermodynamic perturbation method. Units are in kcal/mol. The statistical sampling error of the calculated energetic contribution ($\Delta U$) is $\pm$ 1.0 kcal/mol.}
       \label{TabFF}
    \begin{tabular}{l|c|c|c} 
     \hline
      \textbf{} & \textbf{$\Delta F$} & \textbf{$\Delta U$} & \textbf{-T$\Delta S$}\\
      \hline
      Gas & 0.0 & 0.0 & 0.0 \\
      Surface & -3.4 & -8.2 & 4.8 \\
      Bulk & -1.3 & -8.95 & 7.65 \\
      \hline
    \end{tabular}
  \end{center}
\end{table}

\clearpage

%\bibliography{bib}